\documentclass[aps,prd,10pt,showpacs,amsmath,showkeys,twocolumn,floatfix,amssymb, preprintnumbers, nofootinbib, superscriptaddress]{revtex4}
\usepackage{graphicx}
\usepackage{comment}
\usepackage[usenames]{color}
\usepackage{bm}
\usepackage{ifpdf}
\usepackage{floatrow}
\usepackage{makecell}
\usepackage{caption}

\usepackage[normalem]{ulem}
\usepackage[dvipsnames]{xcolor}
\usepackage[utf8]{inputenc}
\usepackage{hyperref}
\hypersetup{
  pdfnewwindow=true,      
  colorlinks=true,        
  linkcolor=PineGreen,    
  citecolor=PineGreen,    
  filecolor=PineGreen,    
  urlcolor=PineGreen      
}
\newcommand{\be}{\begin{eqnarray}}
\newcommand{\ee}{\end{eqnarray}}

\begin{document}

\title{Anomalous Faraday effect in a $\mathcal{P}\mathcal{T}$-symmetric dielectric slab}

\author{Vladimir~Gasparian}
\email{vgasparyan@csub.edu}

\affiliation{Department of Physics and Engineering,  California State University, Bakersfield, CA 93311, USA}

\author{Peng~Guo}
\email{pguo@csub.edu}

\affiliation{Department of Physics and Engineering,  California State University, Bakersfield, CA 93311, USA}

\author{Esther~J\'odar}
\email{esther.jferrandez@upct.es}

\affiliation{Departamento de F\'isica Aplicada,  Universidad Polit\'ecnica de Cartagena, E-30202 Murcia, Spain}

\date{\today}

\begin{abstract}
 In this letter we discuss a phase transition-like anomalous behavior of Faraday rotation angles in a simple  parity-time ($\mathcal{P}\mathcal{T}$) symmetric   model with two complex $\delta$-potential placed  at both boundaries of a regular dielectric slab.  
In anomalous phase, the value of one of Faraday rotation angles turns   negative, and both angles suffer spectral singularities and yield strong enhancement near singularities. 
\end{abstract}

\maketitle

\paragraph{Introduction: }

Faraday rotation  (FR) is a magneto-optical phenomenon that
rotates the polarization of light. It occurs either due to the internal property
of the medium or  to the external magnetic field  applied.
In  both cases, the dielectric permittivity tensor of the system
becomes anisotropic \cite{lan84}. The Faraday effect  shows a wide range of applications in various fields of modern physics, such as,  (i) measuring magnetic field in astronomy \cite{longair_2011}; (ii) construction of optical isolators for fiber-optic telecommunication systems \cite{doi:10.1063/1.341700};  or (iii) optical circulators that are used in the design of microwave integrated circuits \cite{BERGER2003777,Turner:81,Firby:18}.  Usually it is necessary either a large size or a strong external magnetic field in order to obtain a large FR in bulk magneto-optical materials.
\cite{GARBUSI2003319,doi:10.1063/1.1288509}. 
However,  for small   size systems, where the de Broglie wavelength  is compatible with size of systems,   a
large enhancement of the FR and as well as a change in the sign of the FR can be obtained
  by incorporating several nanoparticles and their
composites in nanomaterials,  see e.g. Refs.~\cite{Uchida_2011,doi:10.1063/1.1625084,PhysRevA.89.023830}.  The aim of this letter is to exhibit that the large enhancement of  Faraday rotation and the anomalous phase transition-like effect may occur when the medium  is parity-time ($\mathcal{P}\mathcal{T}$) symmetric.
Moreover,   in certain range of model parameters of  a  $\mathcal{P}\mathcal{T}$-symmetric system
a {\it non-trivial} transition occurs with a change of sign of FR.   {\it Non-triviality} of the transition of FR  only  happens in a few well-known cases, such as (1)  when the sign of the constant Verdet is changed or when either the magnetic field or
the direction of the light is reversed; (2) in a new type
artificial left handed materials (metamaterials) with negative permittivity $\epsilon$
and permeability $\mu$  in Ref.~\cite{Kaipurath2016,PhysRevLett.85.3966,Yang2016,Caligiuri2016,Cao:13}, the sign of FR is negative because
the refractive index $n$ becomes negative. It should be noted that in metamaterials FR changes the sign only in a narrow frequency range, while in the $\mathcal{P}\mathcal{T}$ system discussed here, the change occurs in a fairly large frequency range.

In recent years,  numerous remarkable novel phenomena have been discovered  within  $\mathcal{P}\mathcal{T}$ symmetric systems    \cite{doi:10.1142/q0178,doi:10.1080/00107500072632,RevModPhys.88.035002,doi:10.1142/S0219887810004816,Mostafazadeh2009,El-Ganainy2018,PhysRevLett.100.103904,PhysRevLett.100.030402}, including real spectra of non-Hermitian operators  \cite{doi:10.1142/q0178,doi:10.1080/00107500072632,doi:10.1142/S0217751X0602951X,Joglekar_2012}, spectral singularities \cite{PhysRevLett.102.220402,Ahmed_2009,PhysRevB.80.165125}, the violation of the normal conservation of the photon flux that  leads to anisotropic transmission resonances \cite{PhysRevA.85.023802}, etc. Most importantly, $\mathcal{P}\mathcal{T}$ symmetric systems have been experimentally  developed in optics \cite{El-Ganainy2018,PhysRevLett.100.103904,PhysRevLett.100.030402},  atomic gases \cite{PhysRevLett.110.083604,Hang:14}, plasmonic waveguides \cite{PhysRevA.89.033829,PhysRevB.89.075136}  or acoustic \cite{PhysRevX.4.031042}.

In this letter, we illustrate that the   Faraday rotation angles of the  polarized light traveling through a $\mathcal{P}\mathcal{T}$-symmetric material displays phase transition-like anomalous behaviors. With a simple $\mathcal{P}\mathcal{T}$-symmetric dielectric slab model, we show that in one phase (normal phase), the angle of Faraday rotation behaves normally as in regular dielectric slab with a positive permittivity, and stay positive all the time as expected.  In the second anomalous phase, the angle of Faraday rotation may change the sign and turn into negative.  Two phases are separated by the parameters of $\mathcal{P}\mathcal{T}$-symmetric model. In addition, the spectral singularities occur in the second anomalous phase. The Faraday rotation angles thus yield a strong enhancement near spectral singularities. In this sense, $\mathcal{P}\mathcal{T}$-systems  seem to be a good candidate for constructing fast tunable and switchable polarization rotational ultrathin magneto-optical devices in a wide frequency range with a giant Faraday rotation.

\begin{figure}
\begin{center}
\includegraphics[width=0.8\textwidth]{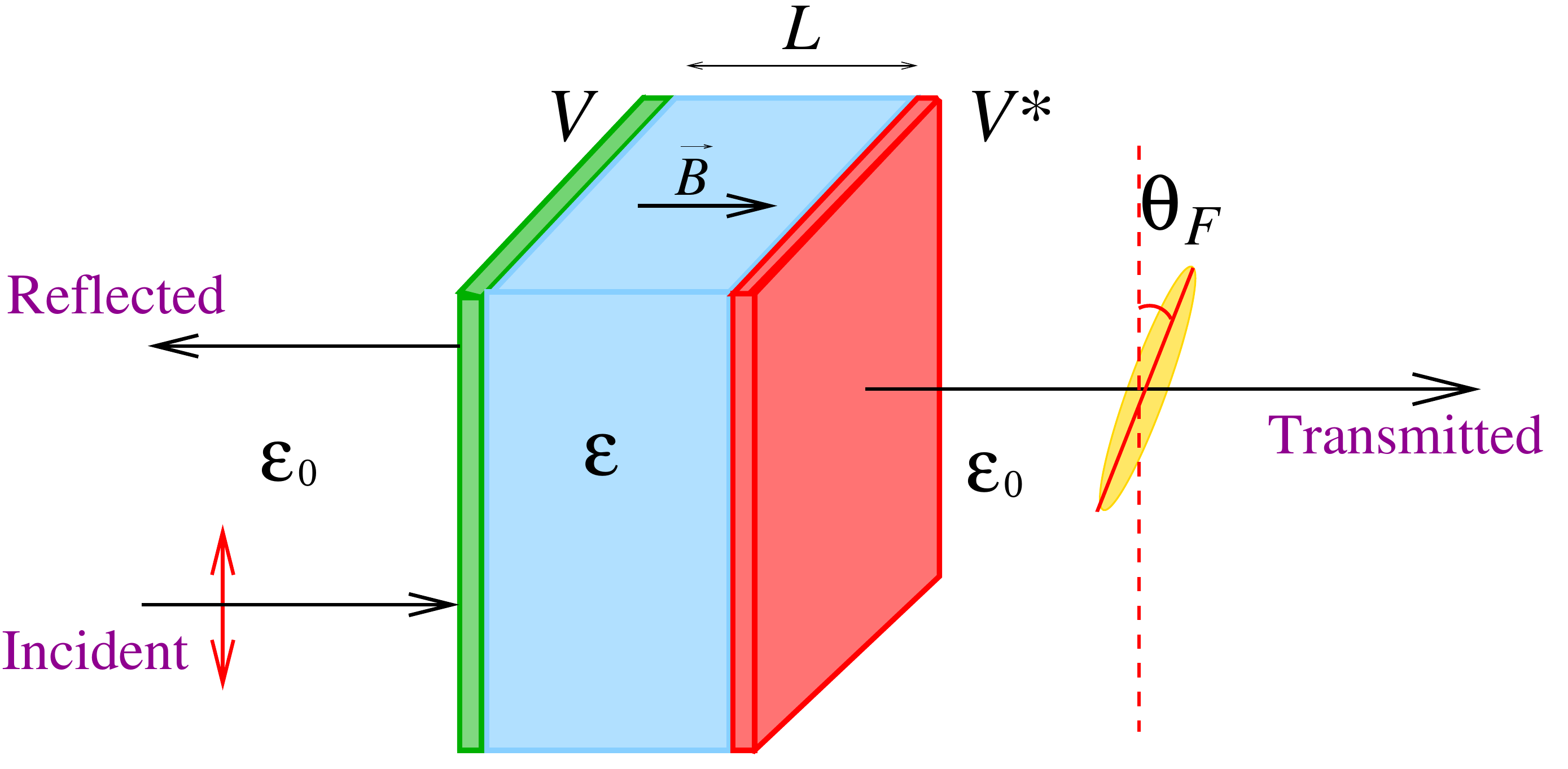}
\caption{ Demo plot of  a $\mathcal{P}\mathcal{T}$-symmetric dielectric slab model with two balanced complex narrow slabs   placed at  both ends of a real dielectric  slab.  }\label{slabplot}
\end{center}
\end{figure}

\paragraph{The theory of Faraday effect in a $\mathcal{P}\mathcal{T}$-symmetric dielectric slab: }

Let us consider a $\mathcal{P}\mathcal{T}$-symmetric dielectric slab  with a  finite spatial extent of length $L$ along $x$ direction, where the permittivity of the slab has balanced gain and loss,
\begin{equation}
    \epsilon (x- \frac{L}{2}) = \epsilon^* (-x + \frac{L}{2}).
\end{equation}
A linearly polarized electromagnetic plane wave with  angular frequency $\omega$  enters the slab from the left at normal incidence propagating along the $x$ direction.  The polarization direction of electric field of incident wave is taken as the z-axis:  $\boldsymbol{ E}_0 (x)=   e^{i k_0 x} \hat{z}$,   where  $k_0=\frac{\omega}{c} \sqrt{\epsilon_0}$ stands for the wave vector and $\epsilon_0$ denotes the dielectric constant of vacuum.  A weak magnetic field $\boldsymbol{ B}$, which preserves the linearity of Maxwell's equations, is applied in the $x$-direction and is confined into the slab, see Fig.~\ref{slabplot}. The scattering of incident wave by the   dielectric slab is described by Schr\"odinger-like equations, see e.g.  Refs.~\cite{PhysRevLett.75.2312,LofyGasparian2020},
\begin{equation}
\left [\frac{d^2}{d x^2}    + \frac{\omega^2 \epsilon_{\pm} (x)}{c^2}  \right ] E_\pm (x)  =0,
\end{equation}
where $E_\pm = E_y \pm i E_z$ are circularly polarized electric fields.   The $\epsilon_{\pm} (x) $ is defined   by, see e.g.  Refs.~\cite{PhysRevLett.75.2312,LofyGasparian2020},
\begin{equation}
\epsilon_{\pm} (x) = 
\begin{cases}  
\epsilon (x) \pm g, \ \ & \ \ x \in [0,L], \\
\epsilon_0, \ \ & \ \ \mbox{otherwise},
 \end{cases}
\end{equation}
 where $g$ is the  gyrotropic vector along the magnetic-field direction. The external magnetic field $\boldsymbol{ B}$ is  included into the gyrotropic vector $g$   to make the calculations
valid for the cases of both external magnetic fields and magneto-optic materials.
The magnetic field causes
the direction of linear polarization to rotate while light
propagates through the medium. As a consequence, 
the electromagnetic wave
is elliptically polarized
and the major axis of the ellipse is rotated with respect
to the original direction of polarization. The angle of Faraday rotation, $\theta_1$, and the degree of ellipticity, $\theta_2$, are defined by \cite{PhysRevLett.75.2312,LofyGasparian2020}
\begin{equation}
\theta_1  = \frac{\psi_+ - \psi_-}{2} , \ \ \ \  \theta_2 = \frac{1}{4} \ln \frac{T_+}{T_-}  ,\label {FR1}
\end{equation}
where $T_\pm$ and $\psi_\pm$ are the transmission coefficients and phase of transmission amplitudes,  $t_\pm=  \sqrt{ T_\pm} e^{i \psi_\pm} $,   of transmitted electric fields: 
\begin{equation}
E_\pm (x > L) = \pm i t_\pm e^{i k_0 x}  .  
\end{equation}

For the dielectric material with real value of permittivity, the scattering $S$-matrix can be parameterized by two independent real scattering phaseshifts, $\delta_\pm^{(1/2)}  $. The transmission coefficients and phases  $\psi_\pm$ are given in terms of scattering phaseshifts by
\begin{equation}
  \sqrt{ T_\pm } = \cos  ( \delta_\pm^{(1)}   - \delta_\pm^{(2)}   ) , \ \ \ \ \psi_\pm =  \delta_\pm^{(1)}   + \delta_\pm^{(2)}    . \label{theta1realpot}
\end{equation}
Hence both angles   $\theta_1 $ and $\theta_2$ are   real and well defined.
As discussed in Ref.~\cite{PhysRevResearch.4.023083}, for the scattering with a complex potential in general,   scattering phaseshifts  become complex. Therefore, with a complex dielectric slab, both   $\theta_1 $ and $\theta_2$    are complex in general, and physical meaning of both angles become ambiguous. 

In a $\mathcal{P}\mathcal{T}$-symmetric system,  the parameterization of  scattering $S$-matrix now requires three  independent real functions: two phaseshifts  and one inelasticity, $\eta_\pm   \in [1,\infty]$, see  Ref.~\cite{PhysRevResearch.4.023083}. The phases of a $\mathcal{P}\mathcal{T}$-symmetric system, $\psi^{(\mathcal{P}\mathcal{T})}_\pm$, are still given by the sum of two real phaseshifts  as in Eq.(\ref{theta1realpot}), thus it remains real. The transmission coefficients of  $\mathcal{P}\mathcal{T}$-symmetric system   also remain real but now depend on inelasticity as well,
\begin{equation}
   \sqrt{ T^{(\mathcal{P}\mathcal{T})}_\pm  } = \eta_\pm   \cos   ( \delta_\pm^{(1)}   - \delta_\pm^{(2)}   )   . \label{Tptetaphase}
\end{equation}
Therefore, with balanced gain and loss, the reality of both  Faraday rotation   angles $\theta_1 $ and $\theta_2$  is warranted in a $\mathcal{P}\mathcal{T}$-symmetric system.  In Ref.~\cite{PhysRevResearch.4.023083}, it is also shown that the generalized Friedel formula relates  the   derivative of sum of two phaseshifts, $\frac{d}{d\omega} (\delta_\pm^{(1)}+\delta_\pm^{(2)})= \frac{d \psi^{(\mathcal{P}\mathcal{T})}_\pm}{d\omega}$,  to the integrated generalized density of states of the $\mathcal{P}\mathcal{T}$-symmetric system, which  turns out to be  real for a $\mathcal{P}\mathcal{T}$-symmetric system. Hence  the reality of FR angles in  a $\mathcal{P}\mathcal{T}$-symmetric system can also be understood based on  the generalized Friedel formula. However, the  positivity of generalized density of state in $\mathcal{P}\mathcal{T}$-symmetric systems is no longer guaranteed. Therefore the FR angles  of $\mathcal{P}\mathcal{T}$-symmetric systems show  an anomalous behavior becoming negative.

\paragraph{A simple $\mathcal{P}\mathcal{T}$-symmetric model: }
 We use in our calculations a very simple $\mathcal{P}\mathcal{T}$-symmetric model to illustrate the anomalous behavior of FR angles by   putting two complex delta potentials   at  both ends of the dielectric  slab with a positive and real   permittivity  ($\epsilon >0$ and real),
\begin{equation}
   \epsilon (x )  = \epsilon + V \delta(x) + V^* \delta(x-L), \ \ \ \ V = V_1 + i V_2 = |V| e^{ i\varphi_V}. \label{doubledeltamodel}
\end{equation}
 We adopt this  model because it lets us obtain quite easily the analytical expressions, and some techniques and conclusions that were developed in   Refs.~\cite{PhysRevLett.75.2312,LofyGasparian2020} can be applied directly in this work. The $\mathcal{P}\mathcal{T}$-symmetric double complex boundaries model is similar to the model by considering $\mathcal{P}\mathcal{T}$-symmetric complex dielectric permittivities of slab: \mbox{$\epsilon(  x \in [0, \frac{L}{2} ]) = \epsilon $} for first half of slab    and \mbox{$\epsilon(  x\in [\frac{L}{2}, L] ) = \epsilon^* $} for another half. Both models can be solved relatively easily and analytically, and both show similar anomalous behaviors of FR angles.   No significant difference between two models have been observed. 
 The basis and physical reason for this conclusion is that the two-boundary model discussed in the manuscript can be considered as having two slabs  with changeable thickness and potential force, so the delta potentials are the limit of the two finite potentials of a square.
 Hence we will simply present some results of double complex boundaries model, and it is sufficient to show anomalous behavior of Faraday rotation angles in  $\mathcal{P}\mathcal{T}$-symmetric systems.

  For weak magnetic field ($g \ll 1$) and constant  dielectric permittivity in the   slab, the  Faraday rotation angles in Eq.(\ref{FR1}) can be evaluated in terms of perturbation expansion of the weak magnetic field. The leading order expressions are obtained in Refs.~\cite{PhysRevLett.75.2312,LofyGasparian2020} and are given by
\begin{equation}
\theta_1    =
 \frac{g}{2n} \frac{\partial \psi}{\partial n} ,  \ \ \ \ 
\theta_{2}  =
 \frac{g}{4 n} \frac{\partial \ln T}{\partial n} ,\label{1mb}
\end{equation}
where
 $n= \sqrt{\epsilon}$ is the refractive index of the slab.
The  $T$ and  ${\psi}$ stand for the
coefficient of transmission and the phase  in the absence of the
external magnetic field $\boldsymbol{ B}$.   
The conclusion in Eq.(\ref{1mb}) also apply to double complex boundaries $\mathcal{P}\mathcal{T}$ model in Eq.(\ref{doubledeltamodel}). For the simple   double complex boundaries $\mathcal{P}\mathcal{T}$-symmetric model, the transmission amplitude, $t= \sqrt{T}e^{i \psi}$,  can be easily obtained by  matching boundary conditions  method. Hence we find
\begin{equation}
    t(\omega) = \frac{e^{-i k_0 L}  \csc (k L) }{R(\omega)- i I(\omega)},
\end{equation}
where
\begin{align}
    R(\omega) &=  \cot (k L) -   Z   \frac{ \omega  V_1  }{c n_0}   , \nonumber \\ .
      I(\omega) & =\frac{\omega  V_1  }{c n_0} \cot (k L)   +    \frac{1  + Z^2\left ( 1  - (  \frac{  \omega  |V|  }{c n_0}    )^2   \right )}{2 Z}       .
\end{align}
$Z=\sqrt{\frac{\epsilon_0}{\epsilon}} = \frac{n_0}{n}$ is the "relative" impedance of the dielectric slab, and   $k=\frac{\omega}{c}n$ denotes the wave vector of propagating waves  inside the dielectric slab. The transmission amplitude $t$ can also be obtained by the Green's function approach \cite{CarpenaGasparianZphysB}. We remark that the Green's function approach is a better tool for more sophisticated multilayer systems,  which is based on the exact calculation
of the Green's function (GF) of a photon for a given dielectric permittivity profile $\epsilon(x)$. The GF approach
is compatible with the transfer matrix method and has been
widely used to calculate the average density of states over a
sample, the energy spectrum of elementary excitations \cite{CarpenaGasparianZphysB}, or
the characteristic barrier tunneling time \cite{PhysRevA.54.4022}, among others.  The coefficient of transmission $T$ and the phase $ \psi$ 
are thus explicitly given by
\begin{equation}  
 T (\omega)   =  \frac{    \csc^2 (k L)   }{    R^2(\omega) + I^2 (\omega)     }    , \ \ \ \ \psi (\omega) = \tan^{-1}\left [ \frac{I (\omega)}{R(\omega)} \right ].
  \end{equation}

 \begin{figure}
\begin{center}
\includegraphics[width=0.8\textwidth]{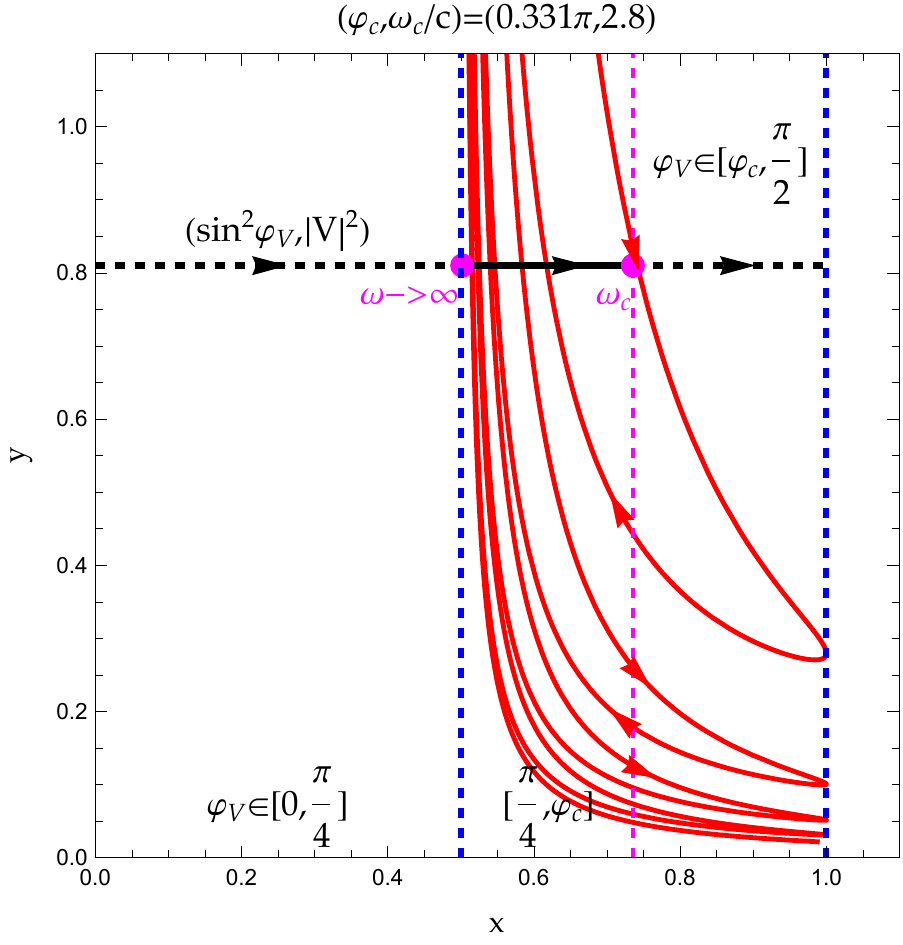}
\caption{ Spectral singularities condition plot: the parametric plot of solid red  curve is generated with $(x,y)$  coordinates    given by left-hand side of  Eq.(\ref{specsingular})     as the function of $\omega/c \in [0,10]$.  The solid red  curve is bound by two blue vertical lines located at $x=\frac{ 1}{ 2  }$ and $x=1$.  The solid black line is generated with coordinates of $(\sin^2 \varphi_V, |V|^2)$ by varying $\varphi_V$ in the range of $[0,\frac{\pi}{4}]$ (dashed black), $[\frac{\pi}{4}, \varphi_c]$ (solid black) and $[\varphi_c,\frac{\pi}{2}]$ (dashed black). The arrows  indicate increasing $\omega$ and $\varphi_V$ directions. The value of  $\omega$   of spectral singularities for fixed $V$ is given by intersection of black line and red curve.    The model parameters are chosen as:  $|V| =0.9 $, $n_0=1$, $L=1$ and $Z= 0.7$.  }\label{specallplot}
\end{center}
\end{figure}

 \begin{figure}
\begin{center}
\includegraphics[width=0.9\textwidth]{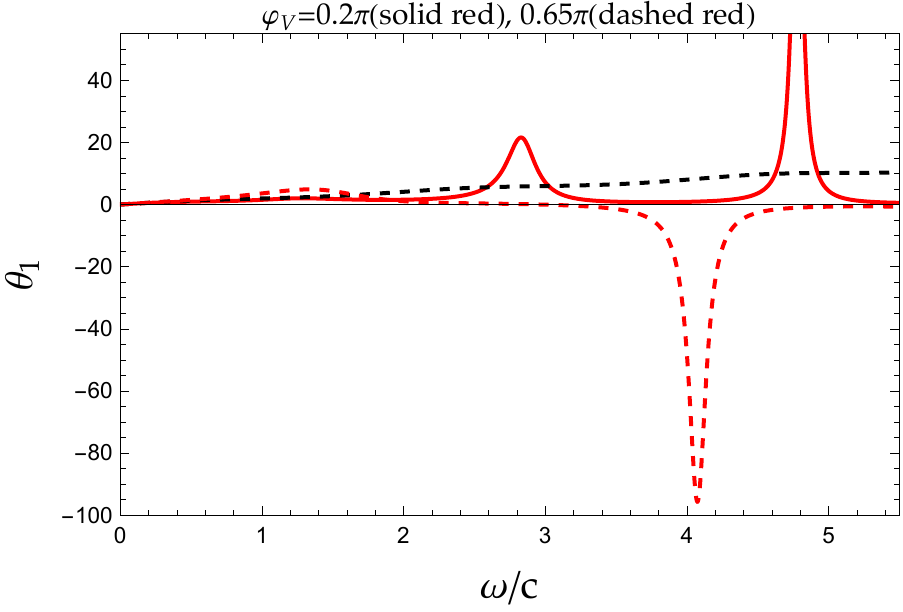}
\caption{  Comparison of   $\theta_1$ of $\mathcal{P}\mathcal{T}$-symmetric model  with various values of $\varphi_V$: $\varphi_V = 0.2 \pi$ (solid red) and $0.65\pi$ (dashed red).  The regular  $\theta_1$ angle (dashed black curve) with no complex boundaries  by setting $V=0$ is also plotted.  
The parameters are taken as:   $|V| =0.9 $, $n_0=1$, $L=1$, and $Z= 0.7$. }\label{theta1plot}
\end{center}
\end{figure}

 \begin{figure}
\begin{center}
\includegraphics[width=0.9\textwidth]{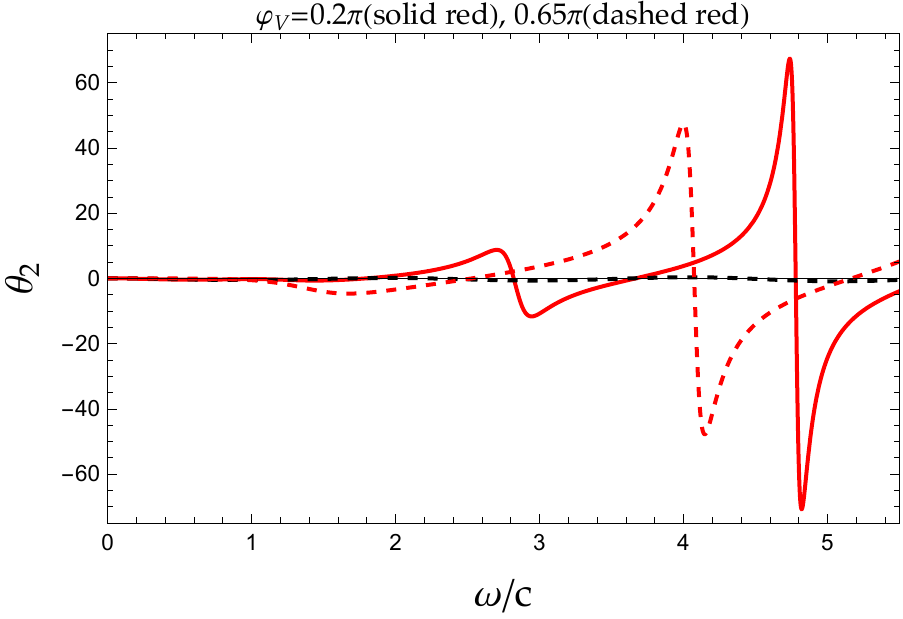}
\caption{  Comparison of   $\theta_2$ of $\mathcal{P}\mathcal{T}$-symmetric model  with various values of $\varphi_V$: $\varphi_V = 0.2 \pi$ (solid red) and $0.65\pi$ (dashed red).  The regular  $\theta_2$ angle (dashed black curve) with no complex boundaries  by setting $V=0$ is also plotted.  
The parameters are taken as:   $|V| =0.9 $, $n_0=1$, $L=1$, and $Z= 0.7$.   }\label{theta2plot}
\end{center}
\end{figure}

We remark that unphysical units are adopted in this work for a simple model: the length of slab $L$ is used to sent up the physical scale,  $V$ and $\epsilon=n^2$ carry the dimensions of $1/L$ and $1/L^2$ respectively. The $\omega/c$ is hence a dimensionless quantity.

\subparagraph{Spectral singularities: }

The resonance states with vanishing spectral width appear in non-Hermitian complex potential scattering theory and yield   divergences of reflection and transmission coefficients of scattered states,  which  are usually referred as spectral singularities, see e.g. Refs.~\cite{PhysRevLett.102.220402,Ahmed_2009,PhysRevB.80.165125}. Therefore, near spectral singularities  it  should be expected a strong enhancement of both Faraday rotation angles. Specifically, for the double complex boundaries $\mathcal{P}\mathcal{T}$-symmetric model considered in this letter,  the spectral singularities occur when both conditions, $R(\omega)=0$ and $I(\omega)=0$, are satisfied.  Hence  the solutions of spectral singularities can be obtained by
  \begin{equation}
     (  \frac{  1   + \frac{1+  \cot^2 (k L)}{Z^2} }{  1  + \frac{1+ 2 \cot^2 (k L) }{Z^2} }, \frac{ 1    + \frac{1+ 2 \cot^2 (k L) }{Z^2}}{  (  \frac{ \omega   }{c n_0}   )^2 } )  =( \sin^2 \varphi_V    ,      |V|^2 )  . \label{specsingular}
  \end{equation}
 Since the left-hand side of the first condition  in Eq.(\ref{specsingular}) is confined  to the range $[\frac{1}{2},1]$,  the spectral singularities exist only when   the phase angle  $\varphi_V$ is     in the range $  [\frac{\pi}{4}, \frac{ \pi}{2}]$ or $  [\frac{\pi}{2}, \frac{3\pi}{4}]$.  Henceforth, all discussions will be   for the range: $\varphi_V \in  [0, \frac{\pi}{2}]$, which is sufficient due to the symmetry of our model.

 The solutions of spectral singularities on real $\omega$ axis can be visualized graphically by observing the intersection of a curve and a line with $(x,y)$ coordinates given by both sides of  Eq.(\ref{specsingular}) for a fixed $|V|$, see Fig.~\ref{specallplot} as a example.  The  curve that is plotted with   coordinates given by left-hand side of  Eq.(\ref{specsingular}) as function of $\omega$ is bound in the region with   $x \in   [\frac{1}{2},1]$. For a fixed $|V|$, the solutions of spectral singularities can only be found in a finite range:  $\varphi_V \in [\frac{\pi}{4}, \varphi_c]$, where $\varphi_c$ stands for upper bound of range, see e.g.  Fig.~\ref{specallplot}.  From Eq.(\ref{specsingular}), the spectral singularity solutions $\omega$ is related to  $\varphi_V$ by $\omega/c = \frac{ n_0}{|V|Z} \frac{1}{\sqrt{|\cos (2\varphi_V) |}}$.  Hence as $\varphi_V$ approaches lower bound of range at $  \frac{\pi}{4}$, the spectral singularity solution occurs at large frequency: $\omega \rightarrow \infty$. When $\varphi_V$ is increased, the solution of  spectral singularity   moves toward lower frequencies.  As $\varphi_V   $ approaches the upper bound of range at $     \varphi_c$,   the spectral singularity solution thus reaches its lowest value at $\omega_c$, see e.g. Fig.~\ref{specallplot}.

\subparagraph{Phase transition-like  phenomenon of Faraday rotation angle $\theta_1$: }
For a regular dielectric slab with positive and real permittivity  ($\epsilon>0$ and real), the Faraday rotation angle $\theta_1$ must be also real and positive. However, in a $\mathcal{P}\mathcal{T}$ system,  the Faraday rotation  angle $\theta_1$ shows a phase transition-like anomalous behavior,   and two phases are separated by model parameter $\varphi_V$: 

(Phase I)  for $\varphi_V \in [ 0,\frac{\pi}{4} ]$,   the value of $\theta_1$ is always positive. No spectral singularities can be found and $\mathcal{P}\mathcal{T}$-symmetric slab behaves just as a regular dielectric slab; 

(Phase II)   for $\varphi_V \in [\frac{\pi}{4},  \frac{\pi}{2}]$,  the value of $\theta_1$ may change the sign and turn negative. When $\varphi_V \in [ \frac{\pi}{4} ,  \varphi_c]$, the spectral singularities occur and   $\theta_1$ starts  showing negative values. The negative   $\theta_1$ only show up at large $\omega$ region when $\varphi_V \sim \frac{\pi}{4}$, and then gradually moves toward the lower frequency region as  $\varphi_V$ is increased. As  $\varphi_V$ continues increasing up to  region of $  [  \varphi_c, \frac{\pi}{2} ]$ that is also free of spectral singularities,    the negativity of  $\theta_1$ persists, see e.g. Fig.~\ref{theta1plot}.

Hence, a $\mathcal{P}\mathcal{T}$ system yields a phase transition-like anomalous behavior of Faraday rotation angle $\theta_1$,  the $\varphi_V= \frac{\pi}{4}$ is the critical value that separates the positivity and negativity phases of $\theta_1$.

The phase transition-like behavior of $\theta_1$ can be understood intuitively by considering limiting case of $|V| \rightarrow \infty$, 
\begin{equation}  
 \theta_1 \stackrel{|V| \rightarrow \infty}{ \rightarrow } \cos (2 \varphi_V )  \frac{g}{2n}   \left(  \frac{ \omega  |V|  }{c n_0} \right)^2  \frac{Z T(\omega)}{2n}   \left (  k L-  \frac{ \sin (2 k L) }{2 }   \right ) \label{theta1Vinfty}.
 \end{equation}
  Now, we can see very clearly that the sign of $\theta_1$ is totally determined by $\varphi_V$ in this limiting case.

  The  anomalous  negativity behavior of $\theta_1$ can also be illustrated analytically at another limiting case by setting $V_1=0$ and $\varphi_V= \frac{\pi}{2}$, 
\begin{equation}  
 \theta_1   \stackrel{V_1 \rightarrow 0}{ \rightarrow } \frac{g T(\omega) }{2n  }    \frac{   \frac{2 k L +\sin (2 k L)}{2}   + Z^2    \frac{2 k L - \sin (2 k L)}{2}     [ 1 - ( \frac{\omega  V_2  }{c n_0})^2  ]   }{2 n_0}. \label{theta1V10}  
 \end{equation}
 Both $2 kL \pm  \sin (2 k L) $ are positive definite functions, hence as $\frac{\omega  V_2  }{c n_0}>1$, the sign of FR angle $\theta_1$ changes from positive to   negative.

\paragraph{Discussion and summary: }

In summary, using a simple $\mathcal{P}\mathcal{T}$-symmetric   model with two complex $\delta$-potential placed  at both boundaries of a regular dielectric slab, we show that FR angles display a phase transition-like anomalous behavior. In regular phase,  Faraday rotation angles behave as normal as in a regular dielectric slab. In anomalous phase, FR angle $\theta_1$ turns negative, and both angles $\theta_1$ and $\theta_2$ suffer spectral singularities and yield strong enhancement near singularities. The critical value of phase transition is controlled by the parameter of $\mathcal{P}\mathcal{T}$-symmetric  model.
A very similar anomalous phase transition-like  behavior to FR is expected   in reflected light (Kerr effect, see e.g. Ref.~\cite{LofyGasparian2020}).

On the contrary to phase transition-like anomalous $\theta_1$ behavior during transition, angle $\theta_2$ doesn't exhibit the significant change of nature except that it also suffers the spectral singularities in anomalous phase.  This can be understood from the definition of $\theta_2$ in Eq.(\ref{FR1}) and the expression of transmission coefficients in Eq.(\ref{Tptetaphase}). The angle  $\theta_2$  is an oscillating function regardless   that dielectric slab is regular or $\mathcal{P}\mathcal{T}$-symmetric.  In $\mathcal{P}\mathcal{T}$ systems, the conservation law of scattering must be generalized, see e.g. Ref.~\cite{PhysRevA.92.062135}, the transmission coefficients are no longer bound in range of $ [0,1]$ due to  the inelasticity functions of $\mathcal{P}\mathcal{T}$ systems  $\eta_\pm \in [1,\infty]$. However, because angle $\theta_2$ depends only on the ratio of $T_\pm$, the oscillating nature of $\theta_2$ remains unchanged even in $\mathcal{P}\mathcal{T}$ systems with unbound $T_\pm$. In addition,   in both phases,    we   observe that $\theta_2$ vanish near frequencies of   resonant peaks appeared in $\theta_1$. The vanishing $\theta_2$ angle  represents the purely  linearly polarized wave with no  
ellipticity. 
The angle $\theta_2$ also displays the sawtooth behavior: the  amplitude of angle $\theta_2$ keep growing  as frequency is increased, and it is periodically repeated and changed sharply both in magnitude and in sign near the resonance frequencies.

 \begin{figure}
\begin{center}
\includegraphics[width=0.9\textwidth]{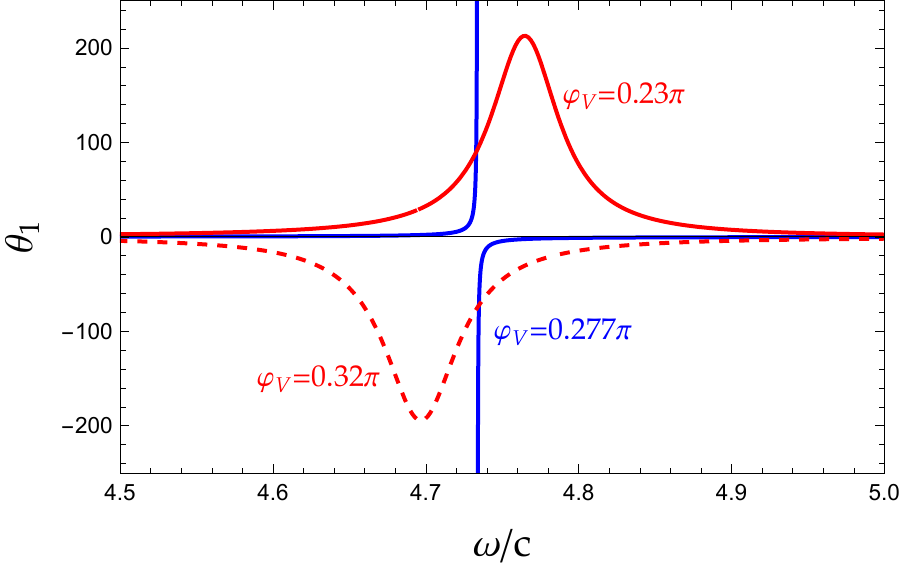}
\caption{    Behavior of  $\theta_1$ of $\mathcal{P}\mathcal{T}$-symmetric model  as   $\varphi_V$ is increased across spectral singularities that is located at $(\omega/c, \varphi_V  ) = (4.73,0.277\pi)$: 
$\varphi_V = 0.23 \pi$ (solid red),  $ \varphi_V = 0.277 \pi$ (solid blue)  and $\varphi_V = 0.32\pi$ (dashed red).  
The parameters are taken as:   $|V| =0.9 $, $n_0=1$, $L=1$, and $Z= 0.7$. }\label{theta1singularityplot}
\end{center}
\end{figure}

As suggested in Ref.~\cite{Guo:2022jyk}, the phase transition-like behavior of $\theta_1$ is closely related to the motion of  the pole of transition amplitudes. Near the pole, the transmission amplitude is approximated by
\begin{equation}
t(\omega) \propto \frac{1}{\omega - \omega_{pole}},
\end{equation}
where $\omega_{pole} = \omega_{Re} + i \omega_{Im}$ is the pole position in complex $k$-plane, see Eq.(24) in \cite{Guo:2022jyk}.  The phase of transmission amplitude is   dominated by  
\begin{equation}
\psi(\omega) \propto \frac{\omega_{Im}}{ (\omega - \omega_{Re})^2 + \omega^2_{Im}}.
\end{equation}
The sign of $\psi(\omega)$ is hence dictated by the position of  the pole. For the regular dielectric slab or in normal phase of $\mathcal{P}\mathcal{T}$ system, the poles remain in unphysical sheet and $\omega_{Im}>0$. However, in anomalous phase of $\mathcal{P}\mathcal{T}$ system,   poles  cross the real axis and move into physical sheet with $\omega_{Im}<0$ and negative $\psi$ near the pole. Therefore the sign of FR angle $\theta_1$ changes as the pole moves across  the real axis from unphysical to physical sheet, and remains negative as long as poles stay in physical sheet.  This   can be easily illustrated in Fig.~\ref{theta1singularityplot}.  For the set of model parameters chosen in Fig.~\ref{theta1singularityplot}, one of the spectral singularities is located at $(\omega/c, \varphi_V  ) = (4.73,0.277\pi)$. For $\varphi_V   $ value slightly below $ 0.277\pi$, the pole is located in unphysical sheet, and $\theta_1$ remains positive. As $\varphi_V   $ value is increased across $ 0.277\pi$, the pole  moves across  the real axis into physical sheet,  and hence $\theta_1$ changes sign and becomes negative.

We remark that for a simple model, the pole indeed yields divergent spectral singularities when  it lies on  the real axis. As discussed in Ref.~\cite{PhysRevA.89.013824}, in realistic systems, the divergence of spectral singularities may be regularized by nonlinearities of systems. Similarly, the regularization of singularities in a realistic system can also be achieved by  imperfection of  $\mathcal{P}\mathcal{T}$ systems with  a slight imbalance between the gain and loss regions by adding an infinitesimal parameter  to imaginary part of the right boundary complex potential.

At last we remark that the reversal of the sign of the Faraday rotation also  occurs in some other special cases. For example, in Ref.~\cite{doi:10.1063/1.3665138}, it was   shown that near the plasmon resonance of $Fe_{2}O_{3}$ nanoparticles solution,   the Faraday rotation exhibits both left and right rotations for fixed frequencies. 
 The latter is due to the change in the sign of the Verdet constant, as  a result of increasing the thickness of the gold shell    with the addition of a gold solution. As also mentioned in Ref.~\cite{doi:10.1063/1.4943651}, the Faraday rotation angle $\theta_1$ can be increased and even changed  its sign using metamaterials to adapt the optical properties of the host system. 
In addition,  for the frequencies range where both complex permittivity $\epsilon$
and permeability $\mu$ have negative non-zero real
parts and positive non-zero imaginary parts, the real part of $n$ turns out to be negative, see, e.g. Ref.~\cite{doi:10.1080/00107510410001667434}. Hence, the angle $\theta_1$, which is an odd function with respect to the refractive index $n$, will change the sign as well.
As for  $\theta_2$, it is an even function of $n$ and does not change sign when $n$ is reversed \cite{LofyGasparian2020}.  The   phase transition-like behavior in the change of sign of the Faraday rotation angle $\theta_1$  in a  $\mathcal{P}\mathcal{T} $   system studied in this letter  demonstrate a quite different mechanism from the cases mentioned above. The phase transition-like anomalous Faraday effect may be observed experimentally for the wider range of frequencies. Hence $\mathcal{P}\mathcal{T}$-systems  seem to be ideal candidates for constructing fast tunable  polarization rotational ultrathin magneto-optical devices.

\acknowledgments

P.G. and V.G.   acknowledge support from the Department of Physics and Engineering, California State University, Bakersfield, CA.  V.G. and E.J. would like to thank UPCT for partial financial support through the concession of "Maria Zambrano ayudas para la recualificación del sistema universitario español 2021-2023" financed by Spanish Ministry of Universities with financial funds "Next Generation" of the EU. We also thank Christopher Wisehart  for improving the use of the English language in the manuscript.

\bibliography{ALL-REF.bib}

\end{document}